\documentclass[runningheads]{llncs}
\usepackage{amsfonts}
\usepackage{amsmath}
\usepackage[T1]{fontenc}
\usepackage{graphicx,verbatim}
\usepackage{booktabs}   
\usepackage{wasysym}    
\usepackage{tikz}       
\usetikzlibrary{arrows.meta}
\begin{document}
\title{Implicit representations are dead. \\Long live explicit primitives!}
%
\author{Nil Stolt-Ans\'o \thanks{Equal contribution}\inst{1} \and
Maik Dannecker\protect\footnotemark[1]\inst{2} \and Wenqi Huang \inst{1} \and \\ Andras Jakab \inst{2} \and Daniel Rueckert \inst{1,3}}
%
\authorrunning{N. Stolt-Ans\'o \& M. Dannecker et al.}
%
\institute{Technical University Munich, Germany \and
University of Zurich, Switzerland
\and
Imperial College London, United Kingdom \\
\email{nil.stolt@tum.de}  --  \email{m.dannecker@tum.de}
}
%

\maketitle              
\begin{abstract}
Continuous parameterization of medical data has emerged as a powerful paradigm for resolution-independent image representation.
While Implicit Neural Representations offer high fidelity and compact storage, their reliance on global Multi-Layer Perceptrons incurs sizeable computational costs, large memory requirements, and extensive optimization times.
As medical imaging trends towards ever-more detailed, high-resolution volumes, these costs impose significant bottlenecks in the applicability of implicit approaches.
Recently, explicit Gaussian-based primitives have revolutionized the representation learning paradigm by trading deep network evaluations for localized, rasterization-friendly primitives.
In this paper, we present a comprehensive, cross-dimensional evaluation of Gaussian representations against implicit approaches for medical imaging applications.
First, we outline a theoretical overview on the mathematical properties offered by explicit primitives beyond what is capable under the implicit neural paradigm.
Subsequently, we benchmark the computational performance on two demanding image datasets: 2D microscopy histology and 3D lung Computed Tomography (CT).
Our experiments demonstrate that Gaussian representations consistently match or surpass reconstruction metrics compared to implicit methods across all compression factors, while displaying significantly lower optimization times, and memory requirements.
Together with the compelling mathematical properties offered by explicit primitives, these findings motivate the wider adoption of Gaussian representations and position them as an attractive direction for future research in medical imaging.
\keywords{Explicit Representations \and Gaussian Primitives \and Implicit Neural Representations \and Medical Image Reconstruction.}
\end{abstract}

\section{Introduction}
Medical images are acquired as discrete voxels of intensities, sampled along structures such as grids.
While standardized, these formats place a limit in spatial and temporal resolution in that a scan's intrinsic detail is directly coupled to its memory footprint.
Off-grid methods instead reframe data as continuous functions, being mostly unaffected by sampling structure and enabling arbitrary upsampling.
Crucially, how this function is parameterized dictates optimization speed, memory dynamics, and ultimately clinical viability.

Traditionally, dictionary learning and sparse coding methods~\cite{aharon2006ksvd,mairal2009online,van2012kernel} represented images as explicit bases, with grid-bound weights dictating the components of the signal at a given location.
With the popularity of deep learning, Implicit Neural Representations (INRs) freed the basis from the grid by storing the signal in the weights of a coordinate Multi-Layer Perceptron (MLP), with expressive activations~\cite{Sitzmann2020SIREN,Saragadam2023WIRE} and encodings~\cite{tancik2020fourierfeatures} overcoming the spectral bias of early ReLU networks.
To speed up the optimization process, latent feature grids~\cite{lombardi2019neural,muller2022instantNGP,chen2022tensorf,fridovich2023kplanes} reintroduced explicit, local feature storage to curb the cost of deep evaluation, retaining only a shallow decoder. 
Most recently, following 3D Gaussian Splatting~\cite{kerbl2023GaussianSplat}, explicit primitive methods~\cite{zhang2024gaussianimage,zhang2025imageGS} abandon neural decoding entirely, representing signals with continuous, freely optimizable atoms.

\begin{figure}[t]
\centering
\includegraphics[trim={0 0.4cm 0 0},width=0.8\textwidth]{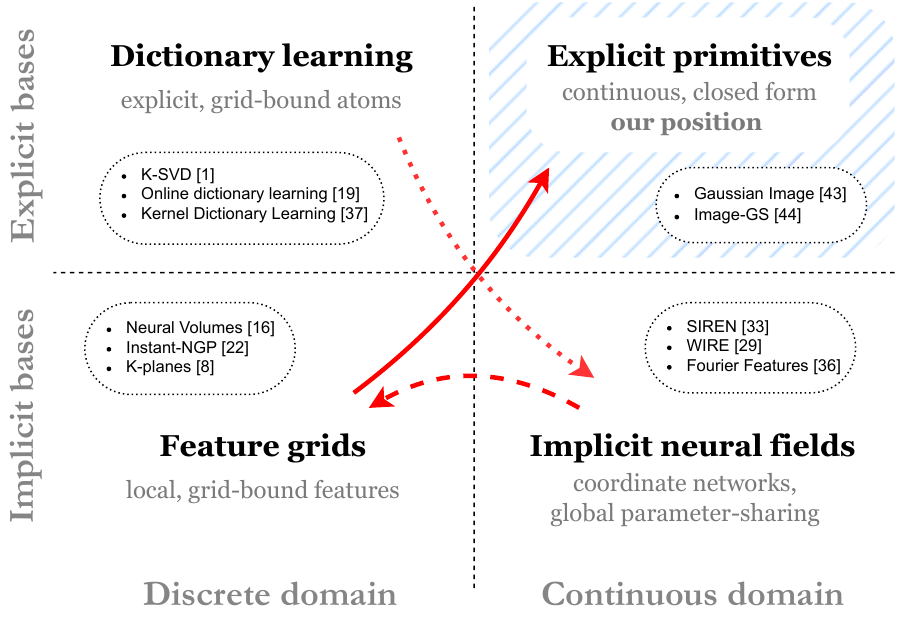}

\caption{Representation paradigms positioned by domain (discrete\,$\to$\,continuous) and basis type (implicit\,$\to$\,explicit). Red arrows indicate the direction of research in recent years: (1-dotted) movement from explicit dictionaries towards neural representations, (2-dashed) shift towards feature-grids, (3-solid) switch to explicit primitives.}
\label{fig:quadrant}
\end{figure}

\subsubsection{Contributions.}
In this work, we argue explicit primitives offer advantages to medical imaging research unmatched by implicit approaches, both in terms of mathematical properties and in computational abilities.  
Concretely: (i) we establish a unified formulation under which dictionaries, INRs, feature grids, and explicit primitives differ only in the parameterization and support of their bases; (ii) we catalogue the closed-form mathematical properties and inductive biases that follow from explicit parametric bases, and show how each unlocks concrete advantages across established and emerging medical imaging applications; and (iii) we benchmark explicit and implicit representations across two real-world medical datasets, showing consistently higher fidelity and robustness, at lower optimization times and memory requirements.

\section{Unified formulation for off-grid methods}
\label{sec:background}
Off-grid methods map a coordinate $\mathbf{x}\in\mathbb{R}^d$ to an intensity, feature, or vector quantity $\mathbf{v}\in\mathbb{R}^c$.
We observe that the dominant paradigms are instances of a single generalized basis expansion,
\begin{equation}
\mathbf{v}(\mathbf{x}) \;=\; \sum_{k\in\mathcal{N}(\mathbf{x})} w_k(\mathbf{x})\,\mathbf{c}_k 
\label{eq:master}
\end{equation}
in which basis (weighting) functions $w_k(\mathbf{x})\in\mathbb{R}$ are combined with coefficients $\mathbf{c}_k\in\mathbb{R}^c$ over an active set $\mathcal{N}(\mathbf{x})\subseteq\{1,\dots,N\}$.
The paradigms differ only in three choices: (i) where the bases $w_k$ originate, (ii) their spatial support together with how the active set $\mathcal{N}(\mathbf{x})$ is selected, and (iii) which quantities are optimized.

\subsubsection{Dictionary methods.}
Sparse coding approximates a signal as $\mathbf{v}\approx\sum_k \alpha_k\,\mathbf{d}_k$, with fixed (learned) atoms $w_k=\mathbf{d}_k$ and scalar coefficients $\mathbf{c}_k=\alpha_k$~\cite{aharon2006ksvd}.
The atoms are discrete vectors defined only at grid samples and the active set $\mathcal{N}$ is chosen \emph{globally} by a sparsity penalty $\lVert\boldsymbol{\alpha}\rVert_0\le T$ rather than by spatial locality.
Only the coefficients (and, in a separate stage, the dictionary) are optimized.
Dictionaries are thus the explicit grid-bound ancestor of the methods below.

\subsubsection{Implicit representations.}
INRs model the signal as a deep MLP $f_{\theta}(\mathbf{x})$ whose final linear layer reads $f_{\theta}(\mathbf{x}) = \sum_k h_k(\mathbf{x})\,p_k$. 
This is exactly Eq.~\eqref{eq:master} where $w_k(\mathbf{x})=h_{k}(\mathbf{x})$ is the penultimate layer's k-th activation and $\mathbf{c}=\mathbf{p}$ are the final layer's weights.
Here, the bases are generated on-the-fly: they have global support ($\mathcal{N}(\mathbf{x})=\{1,\dots,N\}$ for all $\mathbf{x}$), as they are nonlinear functions of \emph{all} weights $\theta$, and optimization targets the generator parameters $\theta$. 
While early ReLU networks exhibited \textit{spectral bias}, periodic~\cite{Sitzmann2020SIREN} and Gabor-wavelet~\cite{Saragadam2023WIRE} activations and Fourier encodings~\cite{tancik2020fourierfeatures} have shown to capture high-frequency expressivity.
Generating bases on demand reduces storage requirements relative to dictionaries, although the MLP must learn to partition the domain through many successive layers, incurring large optimization times for detailed volumes.

\subsubsection{Latent feature grids.}
Feature grids~\cite{lombardi2019neural} take a hybrid approach by introducing explicit local storage alongside an implicit decoder $f_{\theta}$. 
A (multi-resolution) grid of latent feature vectors $\mathbf{Z}$ is instantiated and queried for an individual coordinate via interpolation $\mathbf{Z}(\mathbf{x})\in \mathbb{R}^{F}$.
Feature grids functionally equate to explicitly representing and directly optimizing the intermediate activations $\mathbf{Z}(\mathbf{x})$ of a deep network, bypassing the need for early layers of a network to generate it.
This results in significantly lower depth requirements, considerably diminishing the compute costs of mapping coordinates to bases.
Contextualizing this to fully implicit approaches, the overall function can be summarized as $\mathbf{v}(\mathbf{x}) = f_{\theta}(\mathbf{Z}(\mathbf{x}))$.
Most famously, \cite{muller2022instantNGP} further utilize hardware-specific tricks to improve optimization times.
Alternatively, other approaches factorize dense grids into compact planes and vectors~\cite{chen2022tensorf,fridovich2023kplanes}, alleviating the storage costs of high-resolution feature grids.

\subsubsection{Explicit representations.}
Explicit methods make the bases themselves the free parameters.
For Gaussian primitives, we define the basis as
\begin{equation}
w_k(\mathbf{x}) \;=\; \exp\!\Big(\!-\tfrac{1}{2}\,(\mathbf{x}-\boldsymbol{\mu}_k)^{\!\top}\boldsymbol{\Sigma}_k^{-1}(\mathbf{x}-\boldsymbol{\mu}_k)\Big),
\label{eq:gauss}
\end{equation}
with center $\boldsymbol{\mu}_k\in\mathbb{R}^d$ and anisotropic covariance $\boldsymbol{\Sigma}_k$ parameterized through its Cholesky factors to ensure positive semi-definiteness. 
The Gaussian parameters $\{\boldsymbol{\mu}_k,\boldsymbol{\Sigma}_k\}$, as well as $\mathbf{c}_k$ are optimized.
A Gaussian has infinite support, but its rapid exponential decay renders each primitive negligible beyond a few standard deviations.
Truncating support via k-nearest neighbors further reduces computational requirements with negligible impact on reconstruction metrics.

The additive form of Eq.~\eqref{eq:master} imposes a zero (baseline) prior away from primitives.
Alternatively, the normalized weighting form, known as Partition of Unity (PoU), is more efficient on large uniform regions~\cite{zhang2025imageGS}, albeit forfeiting the analytic closure outlined in Section~\ref{sec:properties}.
This returns to the dictionary principle, yet the atoms are now continuous, may span arbitrarily-large regions, and are freely optimizable. 
Overlapping primitives share bases across neighboring regions, thus scaling capacity with signal complexity rather than grid resolution.

\section{The Explicit Advantage: Analytic, Controllable, and Geometric Representations}
\label{sec:properties}
Section~\ref{sec:background} established that explicit primitives are the only paradigm whose bases are at once continuous, parametric, and geometrically explicit.
We now outline existing and potential applications of these mathematical properties.

\subsection{Analytic closure}
\label{subsec:analytic}


\textbf{Closure under convolution.}
Sensor blur is omnipresent in image acquisition and especially critical in medical imaging, where the measurement is the anatomy convolved with the device's point-spread function (PSF).
Modeling the PSF inside an implicit network demands costly numerical integration, propagating many stochastic samples per query~\cite{NESVOR}.
Since Gaussians are closed under convolution, an explicit Gaussian representation captures the PSF as a simple covariance addition,
$\boldsymbol{\Sigma}_{\text{obs}} = \boldsymbol{\Sigma}_k + \boldsymbol{\Sigma}_{\text{PSF}}$~\cite{dannecker2025fastexplicit}.
The same rasterization extends to tomographic projection, where it already enables sparse-view CT reconstruction an order of magnitude faster than neural fields~\cite{zha2024r2gaussian}.

\textbf{Spatial-spectral duality.}
The Fourier transform of a Gaussian is again a Gaussian, so an explicit representation can be fit directly against raw frequency-domain measurements, bypassing the gridding that discrete Fourier transforms require.
Gabor primitives extend Gaussian primitives by adding frequency and phase components.
Their spectra are origin-shifted Gaussians, further permitting spectral space-partitioning for improved high-frequency recovery~\cite{huang2026gabor}.

\textbf{Analytical Lipschitz continuity.}
Balancing expressiveness and smoothness requires bounding a representation's global upper bound on the function's spatial rate of change, i.e. its Lipschitz constant.
In implicit approaches, this demands complex processes of rigid layer-wise spectral normalizations~\cite{coiffier20241-lipschitz}, regularization techniques~\cite{liu2022learning}, strategic budget distribution across all network components~\cite{mcginnis2026beyond}, or applying smoothing constraints to the output space using control points~\cite{sideri2024sinr}. 
Conversely, the Lipschitz constant of an explicit Gaussian representation is analytically defined directly by covariance matrices $\Sigma_k$ and amplitudes of its primitives, entirely avoiding the architectural overhead and expressiveness bottlenecks inherent to spectrally constrained MLPs.

\textbf{Closed-form derivatives and volumes.}
Spatial derivatives of Eq.~\eqref{eq:gauss} are analytic, and a primitive's volume is proportional to $\sqrt{\lvert\boldsymbol{\Sigma}_k\rvert}$.
While strictly isochoric (volume-preserving) deformations are classically enforced through expensive Jacobian-determinant constraints~\cite{haber2004volume}, Gaussian representations reduce to a simple penalty on $\lvert\boldsymbol{\Sigma}_k\rvert$, sidestepping the double-backpropagation through a neural deformation field.
Explicit Gaussian deformation models already rival iterative and learning-based registration at a fraction of the runtime~\cite{li2024gaussiandir}.

\

\subsection{Controllable optimization}
\label{subsec:control}

\textbf{Structure-aware initialization.}
Random initialization of an MLP is notoriously hard to tune, and data-aware priors demand expensive meta-learning~\cite{tancik2021learnit,sitzmann2020metasdf}.
Explicit primitives can instead be placed deterministically from the target's own structure, concentrating modeling capacity in high-frequency regions before the first gradient step~\cite{zhang2025imageGS}.

\textbf{Adaptive densification.}
Capacity can be allocated where the signal demands it.
New primitives may be spawned at high-error regions during optimization. As their support is effectively local, this resolves fine detail without perturbing already-converged regions or risking catastrophic forgetting~\cite{kerbl2023GaussianSplat}.
Multi-scale primitive densification has proven effective for exactly this in deformable registration~\cite{li2024gaussiandir}.

\textbf{Native coarse-to-fine control.}
Coarse-to-fine optimization is essential to jointly learn representations and spatial transforms, as in pose or motion estimation.
Implicit methods emulate it by annealing the positional encoding~\cite{park2021nerfies}.
Meanwhile, explicit methods obtain it natively by applying a decaying blur to the covariances, granting direct control of the frequencies' upper bound.

\textbf{Inductive biases and extrapolation.}
The blending choice in $w(\mathbf{x})$ is itself a controllable prior.
The additive form decays to a baseline away from primitives, naturally modeling empty uniform pockets (e.g. in the lungs) without the use of any primitives. 
Alternatively, the normalized form (PoU) captures large homogeneous regions~\cite{zhang2025imageGS,dannecker2025fastexplicit} with higher parameter-efficiency.
Explicit primitives further extrapolate predictably beyond the training domain.
This causes bordering background regions to require no additional primitives, deformation and velocity fields stay sensible outside the sampled coordinate range, and permits standardization of cohort volumes without boundary artifacts.

\subsection{Geometric structure}
\label{subsec:geometric}

\textbf{Semantical feature extraction.}
Extracting semantics from an INR is not straight-forward.
Mapping network weights to labels~\cite{deluigi2023inr2vec}, meta-learning segmentation priors~\cite{fitpixels2025}, or auto-decoding a latent vector~\cite{DeepSDF} all scale poorly and tend to overfit intensities at the expense of structure~\cite{NISF}.
Over an explicit primitive set, the representation becomes a point cloud, allowing classification and segmentation to become graph- and node-level problems addressable with permutation-invariant graph networks~\cite{qi2017pointnet,zaheer2017deepsets}.
Message passing over Gaussian primitives already drives classification~\cite{liu2022learning} and segmentation~\cite{jain2024gaussiancut} of splatted scenes, as well as general feature extraction~\cite{qi2025gspr}.

\textbf{Parameter-space regularization}
In implicit neural fields, preventing occupancy in specific regions relies on soft spatial regularizers~\cite{yang2023freenerf,maas2025nerf}.
Because these are output-space constraints, they demand dense numerical integration to penalize the network indirectly.
The transition to explicit parameters offers direct parameter-space intervention.
If an anatomical prior dictates a disallowed region, or if a primitive's contribution falls below a threshold, the parameters can be explicitly pruned or penalized~\cite{kerbl2023GaussianSplat,singh2026backtobasis}.


\subsection{Limitations}
\textbf{Generative modeling.}
Implicit generative models rely on conditioning a multi-layer perceptron (MLP) with a global latent vector $\mathbf{z}$~\cite{DeepSDF}.
Following works increase reconstruction fidelity by anchoring latent codes to specific spatial partitions~\cite{chabra2020deep,mehta2021modulated}. 
Some works try to bridge into the Gaussian domain by showing that Gaussian scenes can be generated by autoregressively predicting latent vectors along a grid, and learning to decode each vector into Gaussian parameters~\cite{vonluetzow2026gaussiangpt}.
Despite the advantages of explicit methods, their off-grid topology introduces a fundamentally conflict with the strict grid-based requirements of current generative architectures.
Although works exist which experiment with utilizing permutation equivariant architectures to generate Gaussian scenes~\cite{shabanov2026free,yushi2025gaussiananything}, these rely ground-truth representations trained on densely acquired data, which is a rare commodity in medical imaging.
Overall, this topic remains severely under-researched, simultaneously opening an exciting frontier for modeling image distributions natively through explicit basis expansions.



\section{Performance evaluation and scaling properties}
Next, we evaluate the performance and scaling properties of explicit representations against common implicit representation techniques.
While representation literature predominantly concerns itself in maximizing reconstruction quality, we argue that the applicability of these methods to real-world research also relies on training requirements. 
As such, we outline two metrics to aid researchers in quantifying hardware bottlenecks for their research:

\textbf{Optimization duration.} 
Optimization time dictates a researcher's ability to perform architectural exploration and hyper-parameter tuning, as well as the size of a cohort for which results may be computed. 
We measure this as the time required for a reconstruction metric to plateau. 
In our experiments, we set the early-stopping criterion as the first 300 training steps elapsing without reconstruction metrics (PSNR) increasing by 0.2dB. 

\textbf{Backpropagation memory usage.}
A key factor influencing the efficacy of training processes is the size of a mini-batch.
The memory capacity of a GPU places a ceiling over the maximum batch size.
This influences the convergence stochasticity and the learning rate that may be employed, and consequently determines the feasibility of learning representations for a given type of data. 
We argue it is important for researchers to be given an up-front understanding on the memory requirements of a given method.
We do so by reporting memory usage per coordinate during the backpropagation of gradients.

\subsubsection{Datasets.}
To reflect applicability of implicit and explicit methods to real-world medical imaging research, we employ two datasets with significantly increased complexity compared to standard datasets used in INR literature.

\textbf{Breast Cancer Histology images (BACH)~\cite{aresta2019bach}.} 
Histology images offer high-resolution detailed color images of complex tissue, containing both regions of high-frequency features and flat uniform regions.
The standard image size is [2048 x 1536], resulting in over 3.1 million RGB samples.

\textbf{National Lung Screening Trial (NLST)~\cite{nlst2011}.}
3D CT chest scans offer complex lung structures as well as intricate acquisition-related noise patterns.
A volume's resolution is [151 x 512 x 512], resulting in over 39.5 million voxels, meaning that representation methods with sensible compression factors will require to be scalable to millions of trainable parameters.

\begin{figure}[!t]
\centering
\includegraphics[trim={0.5cm 0.8cm 0.5cm 0.5cm},width=\textwidth]{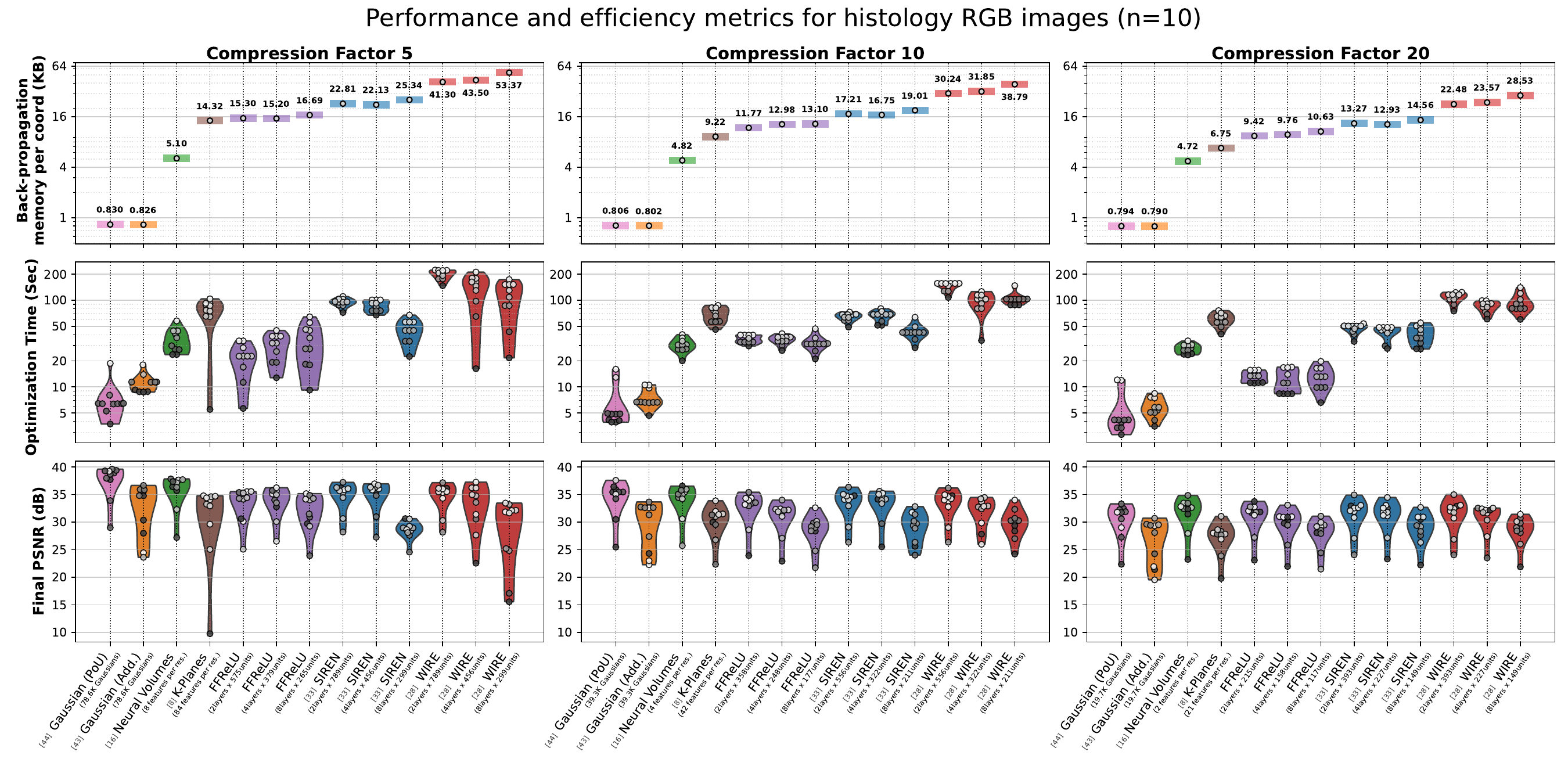}

\caption{Performance and efficiency metrics for various implicit and explicit approaches. Scatter point grayness indicates the number of steps before a run failed to increase PSNR by 0.2 and was early-stopped (black=earlier, white=later).}
\label{fig:BACH}
\end{figure}

\subsubsection{Baselines.}
For all methods, we tuned learning rates via grid-search to determine the training parameter choices for each method.
We observed both datasets having similar optimal learning rate choices and thus use the same for both.
We run all methods for compression factors 5, 10, and 20.
All methods are trained using ADAM optimizer, under a batch size of 200k and mixed-precision floating-point assignments.

\textit{Gaussian methods}. We evaluate two Gaussian weighting schemes: Additive~\cite{zhang2024gaussianimage} and PoU~\cite{zhang2025imageGS}.
We follow the initialization scheme in~\cite{zhang2025imageGS} and use a learning rate of $5 \cdot 10^{-2}$.

\textit{Implicit representations}. We choose three popular fully-implicit baselines: Fourier-features~\cite{tancik2020fourierfeatures}, SIREN~\cite{Sitzmann2020SIREN}, and WIRE~\cite{Saragadam2023WIRE}.
We used a learning rate of $10^{-4}$ for Fourier-features, while SIREN and WIRE appeared to require a learning rate of $10^{-5}$ in order to achieve competitive results.
For Fourier-features we used 256 frequencies under the BACH dataset and 512 for NLST. 
For SIREN a frequency term $\omega$ of 250 performed best for BACH while a lower $\omega$ of 30 performed best for NLST. 
Similarly, WIRE performed best with parameters $\omega = 250$ and $\sigma=0.1$ for BACH, while under NLST WIRE performed best with $\omega=10$ and $\sigma=1$.

\textit{Feature grids}. We include two feature-grid approaches: Multi-resolution Neural Volumes~\cite{lombardi2019neural} and K-planes~\cite{fridovich2023kplanes}.
Both feature grids resolutions start from $16$ and increase by a factor of 2 up until the compression quota is met (resulting in the final resolution being a partial factor of 2).
A learning rate of $10^{-3}$ was used for both Neural Volumes and K-planes.

\subsubsection{Results.}
Gaussian primitives appear to match or surpass implicit baselines in PSNR metrics.
The Gaussian PoU weighting scheme seems to surpass the plain additive scheme, likely to its weighting scheme's improved ability to model uniform regions more efficiently.
While compared to feature grids, Gaussians are 2-5x faster and require 5-15x less memory, the difference grows to 5-100x faster and 10-200x less memory when compared to fully implicit methods.
Fully-implicit methods scale the poorest, requiring the most fitting time and memory, while achieving lower metrics than Gaussian PoU and Neural Volumes. 
Also, as seen in the 8-layer variants in Fig.~\ref{fig:BACH}, fully-implicit methods suffer under increased depths likely due to vanishing gradients. 

\begin{figure}[!t]
\centering
\includegraphics[trim={0.5cm 0.8cm 0.5cm 0.5cm},width=\textwidth]{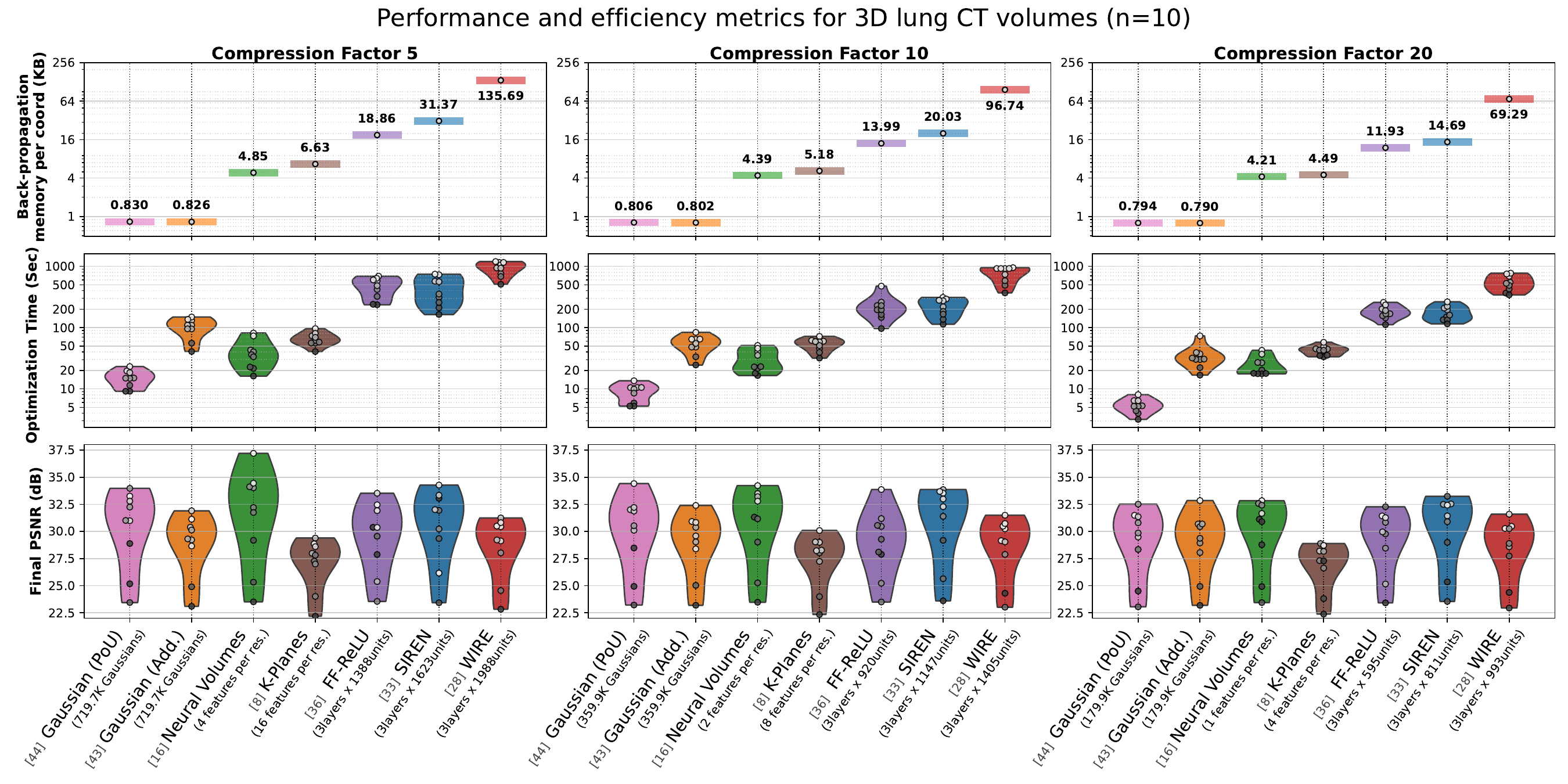}

\caption{Performance and efficiency metrics for various implicit and explicit approaches. Scatter point grayness indicates the number of steps before a run failed to increase PSNR by 0.2 and was early-stopped (black=earlier, white=later).}
\label{fig:NLST}
\end{figure}

\section{Conclusion}
In this work we have consolidated implicit and explicit representations into the same mathematical framework of dictionary methods. 
Furthermore, we have outlined the various mathematical properties that, from a medical imaging perspective, make Gaussian primitives exceptionally interesting over implicit approaches. 
Finally, our empirical results on two datasets show that explicit methods offer the most scalable solution to image representation while matching or surpassing state-of-the-art implicit methods. 
Ultimately, we hope that this work will serve as a catalyst to inspire a new generation of efficient and scalable off-grid paradigm.

%
%
\section*{Acknowledgments}
We would like to thank Julian McGinnis for his endless patience with our continuous Gaussian discussions despite his implicit preferences. 
This work is funded by the Munich Center for Machine Learning.

\section*{Disclosure of Interests}
The authors have no competing interests to declare that are relevant to the content of this article.

{
\footnotesize 
\bibliographystyle{splncs04}
\bibliography{references_compressed}

@article{aharon2006ksvd,
  title   = {{K-SVD}: An algorithm for designing overcomplete dictionaries for sparse representation},
  author  = {Aharon, Michal and Elad, Michael and Bruckstein, Alfred},
  journal = {IEEE Trans. Signal Process.}, volume = {54}, number = {11}, pages = {4311--4322}, year = {2006}
}

@inproceedings{mairal2009online,
  title     = {Online dictionary learning for sparse coding},
  author    = {Mairal, Julien and Bach, Francis and Ponce, Jean and Sapiro, Guillermo},
  booktitle = {ICML}, year = {2009}
}

@inproceedings{van2012kernel,
  title     = {Kernel dictionary learning},
  author    = {Van Nguyen, Hien and Patel, Vishal M. and Nasrabadi, Nasser M. and Chellappa, Rama},
  booktitle = {ICASSP}, year = {2012}
}

@inproceedings{Sitzmann2020SIREN,
  title     = {Implicit neural representations with periodic activation functions},
  author    = {Sitzmann, Vincent and Martel, Julien N.P. and Bergman, Alexander W. and Lindell, David B. and Wetzstein, Gordon},
  booktitle = {NeurIPS}, year = {2020}
}

@inproceedings{Saragadam2023WIRE,
  title     = {{WIRE}: Wavelet implicit neural representations},
  author    = {Saragadam, Vishwanath and LeJeune, Daniel and Tan, Jasper and Balakrishnan, Guha and Veeraraghavan, Ashok and Baraniuk, Richard G.},
  booktitle = {CVPR}, year = {2023}
}

@inproceedings{tancik2020fourierfeatures,
  title     = {Fourier features let networks learn high frequency functions in low dimensional domains},
  author    = {Tancik, Matthew and Srinivasan, Pratul P. and Mildenhall, Ben and Fridovich-Keil, Sara and Raghavan, Nithin and Singhal, Utkarsh and Ramamoorthi, Ravi and Barron, Jonathan T. and Ng, Ren},
  booktitle = {NeurIPS}, year = {2020}
}

@article{lombardi2019neural,
  title   = {Neural volumes: Learning dynamic renderable volumes from images},
  author  = {Lombardi, Stephen and Simon, Tomas and Saragih, Jason and Schwartz, Gabriel and Lehrmann, Andreas and Sheikh, Yaser},
  journal = {arXiv preprint arXiv:1906.07751}, year = {2019}
}

@article{muller2022instantNGP,
  title   = {Instant neural graphics primitives with a multiresolution hash encoding},
  author  = {M\"{u}ller, Thomas and Evans, Alex and Schied, Christoph and Keller, Alexander},
  journal = {ACM Trans. Graph.}, volume = {41}, number = {4}, pages = {1--15}, year = {2022}
}

@inproceedings{chen2022tensorf,
  title     = {{TensoRF}: Tensorial radiance fields},
  author    = {Chen, Anpei and Xu, Zexiang and Geiger, Andreas and Yu, Jingyi and Su, Hao},
  booktitle = {ECCV}, year = {2022}
}

@inproceedings{fridovich2023kplanes,
  title     = {{K-Planes}: Explicit radiance fields in space, time, and appearance},
  author    = {Fridovich-Keil, Sara and Meanti, Giacomo and Warburg, Frederik Rahb{\ae}k and Recht, Benjamin and Kanazawa, Angjoo},
  booktitle = {CVPR}, year = {2023}
}

@article{kerbl2023GaussianSplat,
  title   = {3D {Gaussian} splatting for real-time radiance field rendering},
  author  = {Kerbl, Bernhard and Kopanas, Georgios and Leimk\"{u}hler, Thomas and Drettakis, George},
  journal = {ACM Trans. Graph.}, volume = {42}, number = {4}, pages = {1--14}, year = {2023}
}

@inproceedings{zhang2024gaussianimage,
  title     = {{GaussianImage}: 1000 {FPS} image representation and compression by 2D {Gaussian} splatting},
  author    = {Zhang, Xinjie and Ge, Xingtong and Xu, Tongda and He, Dailan and Wang, Yan and Qin, Hongwei and Lu, Guo and Geng, Jing and Zhang, Jun},
  booktitle = {ECCV}, year = {2024}
}

@inproceedings{zhang2025imageGS,
  title     = {{Image-GS}: Content-adaptive image representation via 2D {Gaussians}},
  author    = {Zhang, Yunxiang and Li, Bingxuan and Kuznetsov, Alexandr and Jindal, Akshay and Diolatzis, Stavros and Chen, Kenneth and Sochenov, Anton and Kaplanyan, Anton and Sun, Qi},
  booktitle = {ACM SIGGRAPH}, year = {2025}
}

@inproceedings{park2021nerfies,
  title     = {Nerfies: Deformable neural radiance fields},
  author    = {Park, Keunhong and Sinha, Utkarsh and Barron, Jonathan T. and Bouaziz, Sofien and Goldman, Dan B. and Seitz, Steven M. and Martin-Brualla, Ricardo},
  booktitle = {ICCV}, year = {2021}
}

@article{NESVOR,
  title   = {{NeSVoR}: Implicit neural representation for slice-to-volume reconstruction in {MRI}},
  author  = {Xu, Junshen and Moyer, Daniel and Gagoski, Borjan and Iglesias, Juan Eugenio and Grant, P. Ellen and Golland, Polina and Adalsteinsson, Elfar},
  journal = {IEEE Trans. Med. Imaging}, volume = {42}, number = {6}, pages = {1707--1719}, year = {2023}
}

@article{dannecker2025fastexplicit,
  title   = {Fast and explicit: Slice-to-volume reconstruction via 3D {Gaussian} primitives with analytic point spread function modeling},
  author  = {Dannecker, Maik and others},
  journal = {arXiv preprint arXiv:2512.11624}, year = {2025}
}

@article{huang2026gabor,
  title   = {{Gabor} primitives for accelerated cardiac cine {MRI} reconstruction},
  author  = {Huang, Wenqi and others},
  journal = {arXiv preprint arXiv:2603.05681}, year = {2026}
}

@inproceedings{singh2026backtobasis,
  title     = {Back to Basis: Spatially continuous {MRI} with adaptive {Gaussians}},
  author    = {Singh, Imraj and Dupuis, Andrew and Kukran, Simran and Badve, Chaitra and Griswold, Mark},
  booktitle = {ISMRM Workshop on Data Sampling and Image Reconstruction}, year = {2026}
}

@inproceedings{mcginnis2026beyond,
  title     = {Beyond uniformity: Regularizing implicit neural representations through a {Lipschitz} lens},
  author    = {McGinnis, Julian and Shit, Suprosanna and H{\"o}lzl, Florian A. and Friedrich, Paul and B{\"u}schl, Paul and Sideri-Lampretsa, Vasiliki and M{\"u}hlau, Mark and Cattin, Philippe C. and Menze, Bjoern and Rueckert, Daniel and others},
  booktitle = {ICLR}, year = {2026}
}

@article{coiffier20241-lipschitz,
  title   = {1-{Lipschitz} neural distance fields},
  author  = {Coiffier, Guillaume and B{\'e}thune, Louis},
  journal = {Comput. Graph. Forum}, volume = {43}, number = {5}, pages = {e15128}, year = {2024}
}

@inproceedings{liu2022learning,
  title     = {Learning smooth neural functions via {Lipschitz} regularization},
  author    = {Liu, Hsueh-Ti Derek and Williams, Francis and Jacobson, Alec and Fidler, Sanja and Litany, Or},
  booktitle = {ACM SIGGRAPH}, year = {2022}
}

@inproceedings{sideri2024sinr,
  title     = {{SINR}: Spline-enhanced implicit neural representation for multi-modal registration},
  author    = {Sideri-Lampretsa, Vasiliki and McGinnis, Julian and Qiu, Huaqi and Paschali, Magdalini and Simson, Walter and Rueckert, Daniel},
  booktitle = {MIDL}, year = {2024}
}

@inproceedings{tancik2021learnit,
  title     = {Learned initializations for optimizing coordinate-based neural representations},
  author    = {Tancik, Matthew and Mildenhall, Ben and Wang, Terrance and Schmidt, Divi and Srinivasan, Pratul P. and Barron, Jonathan T. and Ng, Ren},
  booktitle = {CVPR}, year = {2021}
}

@inproceedings{sitzmann2020metasdf,
  title     = {{MetaSDF}: Meta-learning signed distance functions},
  author    = {Sitzmann, Vincent and Chan, Eric R. and Tucker, Richard and Snavely, Noah and Wetzstein, Gordon},
  booktitle = {NeurIPS}, year = {2020}
}

@inproceedings{deluigi2023inr2vec,
  title     = {Deep learning on implicit neural representations of shapes},
  author    = {De Luigi, Luca and Cardace, Adriano and Spezialetti, Riccardo and Zama Ramirez, Pierluigi and Salti, Samuele and Di Stefano, Luigi},
  booktitle = {ICLR}, year = {2023}
}

@inproceedings{fitpixels2025,
  title     = {Fit pixels, get labels: Meta-learned implicit networks for image segmentation},
  author    = {Vyas, Kushal and Veeraraghavan, Ashok and Balakrishnan, Guha},
  booktitle = {MICCAI}, year = {2025}
}

@inproceedings{mehta2021modulated,
  title     = {Modulated periodic activations for generalizable local functional representations},
  author    = {Mehta, Ishit and Gharbi, Micha{\"e}l and Barnes, Connelly and Shechtman, Eli and Ramamoorthi, Ravi and Chandraker, Manmohan},
  booktitle = {ICCV}, year = {2021}
}

@inproceedings{chabra2020deep,
  title     = {Deep local shapes: Learning local {SDF} priors for detailed 3D reconstruction},
  author    = {Chabra, Rohan and Lenssen, Jan E. and Ilg, Eddy and Schmidt, Tanner and Straub, Julian and Lovegrove, Steven and Newcombe, Richard},
  booktitle = {ECCV}, year = {2020}
}

@inproceedings{DeepSDF,
  title     = {{DeepSDF}: Learning continuous signed distance functions for shape representation},
  author    = {Park, Jeong Joon and Florence, Peter and Straub, Julian and Newcombe, Richard and Lovegrove, Steven},
  booktitle = {CVPR}, year = {2019}
}

@inproceedings{NISF,
  title     = {{NISF}: Neural implicit segmentation functions},
  author    = {Stolt-Ans{\'o}, Nil and McGinnis, Julian and Pan, Jiazhen and Hammernik, Kerstin and Rueckert, Daniel},
  booktitle = {MICCAI}, year = {2023}
}

@inproceedings{zha2024r2gaussian,
  title     = {{R$^2$-Gaussian}: Rectifying radiative {Gaussian} splatting for tomographic reconstruction},
  author    = {Zha, Ruyi and Lin, Tao Jun and Cai, Yuanhao and Cao, Jiwen and Zhang, Yanhao and Li, Hongdong},
  booktitle = {NeurIPS}, year = {2024}
}

@article{li2024gaussiandir,
  title   = {Gaussian primitives for deformable image registration},
  author  = {Li, Xia and others},
  journal = {arXiv preprint arXiv:2406.03394}, year = {2024}
}

@article{haber2004volume,
  title   = {Numerical methods for volume preserving image registration},
  author  = {Haber, Eldad and Modersitzki, Jan},
  journal = {Inverse Probl.}, volume = {20}, number = {5}, pages = {1621--1638}, year = {2004}
}

@inproceedings{qi2017pointnet,
  title     = {{PointNet}: Deep learning on point sets for 3D classification and segmentation},
  author    = {Qi, Charles R. and Su, Hao and Mo, Kaichun and Guibas, Leonidas J.},
  booktitle = {CVPR}, year = {2017}
}

@inproceedings{zaheer2017deepsets,
  title     = {Deep sets},
  author    = {Zaheer, Manzil and Kottur, Satwik and Ravanbakhsh, Siamak and P{\'o}czos, Barnab{\'a}s and Salakhutdinov, Ruslan and Smola, Alexander J.},
  booktitle = {NeurIPS}, year = {2017}
}

@inproceedings{jain2024gaussiancut,
  title     = {{GaussianCut}: Interactive segmentation via graph cut for 3D {Gaussian} splatting},
  author    = {Jain, Umangi and Mirzaei, Ashkan and Gilitschenski, Igor},
  booktitle = {NeurIPS}, year = {2024}
}

@inproceedings{qi2025gspr,
  title     = {{GSPR}: Multimodal place recognition using 3D {Gaussian} splatting for autonomous driving},
  author    = {Qi, Zhangshuo and Ma, Junyi and Xu, Jingyi and Zhou, Zijie and Cheng, Luqi and Xiong, Guangming},
  booktitle = {IROS}, year = {2025}
}

@article{maas2025nerf,
  title   = {{NeRF-CA}: Dynamic reconstruction of X-ray coronary angiography with extremely sparse views},
  author  = {Maas, Kirsten W.H. and Ruijters, Danny and Vilanova, Anna and Pezzotti, Nicola},
  journal = {IEEE Trans. Vis. Comput. Graph.}, year = {2025}
}

@inproceedings{yang2023freenerf,
  title     = {{FreeNeRF}: Improving few-shot neural rendering with free frequency regularization},
  author    = {Yang, Jiawei and Pavone, Marco and Wang, Yue},
  booktitle = {CVPR}, year = {2023}
}

@article{aresta2019bach,
  title   = {{BACH}: Grand challenge on breast cancer histology images},
  author  = {Aresta, Guilherme and others},
  journal = {Med. Image Anal.}, volume = {56}, pages = {122--139}, year = {2019}
}

@article{nlst2011,
  title   = {The National Lung Screening Trial: Overview and study design},
  author  = {{National Lung Screening Trial Research Team}},
  journal = {Radiology}, volume = {258}, number = {1}, pages = {243--253}, year = {2011}
}

@inproceedings{yushi2025gaussiananything,
  title     = {{GaussianAnything}: Interactive point cloud flow matching for 3D generation},
  author    = {Lan, Yushi and Zhou, Shangchen and Lyu, Zhaoyang and Hong, Fangzhou and Yang, Shuai and Dai, Bo and Pan, Xingang and Loy, Chen Change},
  booktitle = {ICLR}, year = {2025}
}

@article{vonluetzow2026gaussiangpt,
  title   = {{GaussianGPT}: Towards autoregressive 3D {Gaussian} scene generation},
  author  = {von L{\"u}tzow, Nicolas and R{\"o}{\ss}le, Barbara and Schmid, Katharina and Nie{\ss}ner, Matthias},
  journal = {arXiv preprint arXiv:2603.26661}, year = {2026}
}

@article{shabanov2026free,
  title={Free-Range Gaussians: Non-Grid-Aligned Generative 3D Gaussian Reconstruction},
  author={Shabanov, Ahan and Hedman, Peter and Weber, Ethan and Li, Zhengqin and Rozumny, Denis and Lan, Gael Le and Dhingra, Naina and Luo, Lei and Vedaldi, Andrea and Richardt, Christian and others},
  journal={arXiv preprint arXiv:2604.04874},
  year={2026}
}
}
\end{document}